\documentclass[%
twoside=semi,%
DIV=13,%
numbers=endperiod,%
headsepline=true,%
headinclude=false,%
footinclude=false,%
fontsize=13pt]%
{scrartcl}

\usepackage[T1]{fontenc}
\usepackage{graphicx}
\usepackage{textcomp}
\usepackage{amssymb}
\usepackage{color}             
\usepackage{url}        
\usepackage{times}
\usepackage{mathptm}
\usepackage{scrpage2}
\usepackage[round]{natbib}
\usepackage{hyperref}

\makeatletter
\newcommand*{\affiliation}[1]{\gdef\@affiliation{#1}}
\renewcommand*{\maketitle}{%
   {\raggedright
   \rmfamily
   {\bfseries%
   \large
   \@title\\[4ex]
   {\small\@author\\[4ex]}}
   \ifdefined\@affiliation%
     {\small\@affiliation\\[6ex]}%
   \fi
   \ifdefined\@date
     {\normalfont\small\@date}
   \fi
   \vspace{4ex}
   }}
\makeatother

\renewenvironment*{abstract}%
{\noindent \textbf{Abstract\,}}%
{\par\noindent}

\setkomafont{section}{\rmfamily\normalsize\bfseries}
\setkomafont{subsection}{\normalfont\rmfamily\normalsize}
\setkomafont{subsubsection}{\normalfont\rmfamily\normalsize\textit}
\setkomafont{captionlabel}{\footnotesize\bfseries}
\setkomafont{caption}{\normalfont\footnotesize}
\setcapindent{0pt}

\newcommand{\frep}{\ensuremath{f_\mathrm{rep}}}
\newcommand{\fceo}{\ensuremath{f_\mathrm{0}}}

\begin{document}

\thispagestyle{empty}
\title{A Laser Frequency Comb System for Absolute\\ Calibration of the VTT
    Echelle Spectrograph}
\author{%
    H.-P.~Doerr$^{1}$\,\textbullet\,
    T.~Steinmetz$^{2,3}$\,\textbullet\,
    R.~Holzwarth$^{2,3}$\\
    T.~Kentischer$^{1}$\,\textbullet\,
    W.~Schmidt$^{1}$
}
\affiliation{%
    $^{1}$ Kiepenheuer-Institut f\"ur Sonnenphysik, Freiburg, Germany\\[-.6ex]
    $^{2}$ Max-Planck-Institut f\"ur Quantenoptik, Garching, Germany\\[-.6ex]
    $^{3}$ Menlo Systems GmbH, Martinsried, Germany%
}
  
  \date{Submitted to SolarPhysics: 30 November 2011, Accepted: 24 February 2012. Published Online: 28 March 2012\\[1ex]
  This reprint was typeset by the Author with permission by Springer/SolarPhysics.\\
  The final publication is available at:\\ \url{http://www.springerlink.com/content/3j087q0121n770h1}\\
  	DOI 10.1007/s11207-012-9960-5}

  
  \maketitle
    
\begin{abstract}
A wavelength calibration system based on a laser frequency comb (LFC) was
developed in a co-operation between the Kiepenheuer-Institut f\"ur Sonnenphysik, Freiburg,
Germany and the Max-Planck-Institut f\"ur Quantenoptik, Garching, Germany for permanent
installation at the German \textit{Vacuum Tower Telescope} (VTT) on Tenerife, Canary Islands.
The system was installed successfully in October 2011. By simultaneously recording the
spectra from the Sun and the LFC, for each exposure a calibration curve can be derived
from the known frequencies of the comb modes that is suitable for absolute calibration at
the meters per second level. We briefly summarize some topics in solar physics that benefit
from absolute spectroscopy and point out the advantages of LFC compared to traditional
calibration techniques. We also sketch the basic setup of the VTT calibration system and its
integration with the existing echelle spectrograph.
\end{abstract}

\section{Introduction}

\subsection{The Need for High-Accuracy Absolute Calibration in Solar Spectroscopy}

The wavelength of spectral lines formed on the solar photosphere is influenced by many
well-known processes. Some of the observed wavelength shifts are due to a true Doppler
shift introduced by motion of the plasma on the solar photospheric surface, e.g., rotation
(2000\,m\,s\textsuperscript{-1}), granulation (500\,m\,s\textsuperscript{-1}), and solar oscillations (500\,m\,s\textsuperscript{-1}).
The convective blueshift and gravitational redshift introduce wavelength shifts on the order of 500\,m\,s\textsuperscript{-1},
while magnetic fields alter the magnitude of the convective blueshift itself, but these effects may not be interpreted as Doppler shifts. Also, the rapidly changing line-of-sight velocity between an observer on the Earth and the observed feature on the Sun amounts to
roughly 1000\,m\,s\textsuperscript{-1}.

To measure small-scale velocities of the plasma on the solar surface, all these effects
must be separated carefully. Sensitive and stable instruments are needed to measure Doppler
velocities of a few meters per second on top of the solar ``noise'' because long averaging
times are required.

While most processes can be characterized with spectroscopic measurements that are
stable to an arbitrarily chosen zero point, precise absolute calibration is crucial to some
measurements of small-scale Doppler velocities. We shall now summarize some of the topics
in solar physics that would benefit from high-precision absolute wavelength measurements.

\subsubsection{Convective Blueshift and Limb Effect}

The convective blueshift is a shift of spectral lines of up to 500\,m\,s\textsuperscript{-1} caused by solar
granulation  \citep{BecklersNelson1978}. The blueshift is strongest at the disk center and shows
a center-to-limb variation, which is also known as the ``limb effect''  \citep[see, e.g.][]{Balthasar1984}.
 The limb effect further complicates Doppler measurements on the solar surface, as
it cannot be averaged out by taking longer time series. Instead, a precise characterization of
the limb effect depending on the position on the solar disk and the spectral line is required
to correct measurements. However, so far, absolute measurements of the limb function have
been limited to a few lines where accurate wavelength references are available
 \citep[e.g.][]{KentischerAndSchroeter1991} and model computations are employed for calibration purposes
\citep{deLaCruzRodriguez.etal.2011}. For any precise spectroscopic measurement of the meridional flow as described below, a preceding determination of the limb
function is a prerequisite, as the effect could also vary with the solar activity.

\subsubsection{Meridional  Motion}

The meridional motion is a weak, poleward directed flow on the solar surface with an am-
plitude of about 20\,m\,s\textsuperscript{-1}. At high latitudes, the flow is expected to sink to deeper layers and
turn equatorward, resulting in a circulation whose revolution time scale might be connected
to the 11-year solar magnetic cycle \citep[see, e.g.][]{Dikpati.and.Gilman.2007}. So far, most mea-
surements of the meridional flow are derived from feature tracking or helioseismic methods.
However, the velocity derived from feature tracking is different from the true plasma mo-
tions, and helioseismology does not give reliable results at the surface due to noise from
solar granulation. Thus, Doppler velocities derived from absolute calibrated high-resolution
spectra would complement the existing measurements.

\subsubsection{Sun-as-a-Star Radial Velocity vs. Solar Activity}

Stellar radial velocity measurements have now reached a precision well below the meters
per second scale  \citep[eg.][]{lovis.etal.2008}, and the required 5\,cm\,s\textsuperscript{-1}
sensitivity to detect Earth-like planets orbiting Sun-like stars seems to be feasible. In this radial velocity regime,
Doppler noise from stellar activity usually dominates the detection limit, as it readily exceeds the Doppler signal from small orbiting planets and the instrument stability. Studying how the solar activity affects the radial velocity signal of the Sun thus is of great importance in detecting signatures of Earth-like exoplanets orbiting Sun-like stars. Detailed
studies were readily carried out on this subject \citep[e.g.][]{lagrange2009,Makarov2010},
 and high-resolution solar spectra with an absolute wavelength calibration
will be useful to verify model calculations and measurements taken with different instruments.

\subsection{Wavelength Calibration Techniques}

Several traditional calibration techniques commonly used in astronomy provide reference
lines that are stable at the meters per second level. However, for absolute wavelength calibration, stability of the used reference lines is necessary but not sufficient, as their absolute
laboratory wavelength must also be known with a corresponding accuracy.

The ideal wavelength reference for spectrograph calibration provides a dense spectrum of
equidistant emission lines of equal intensity with a line width smaller than the instrument's
resolution. The frequencies of the reference lines should be derivable from fundamental
physics so that the calibration source becomes exchangeable without affecting the calibration.

Different calibration techniques were compared with frequency comb calibrators by
 \citet{Murphy2007} for their suitability for extremely stable night-time spectrograph calibration. They concluded that so far only frequency combs allow one to exploit the limits of
existing and future high-precision spectrographs. While most of their findings are also applicable for solar spectroscopy, some aspects are different in this field.

In night-time spectroscopy, thorium-argon lamps are commonly used for calibration. The
Th-Ar spectrum covers the visible spectrum with thousands of sharp lines, many of which
are known with an accuracy of about 10\,m\,s\textsuperscript{-1} \citep{LovisAndPepe2007}. But while the dim
intensity of these lamps is sufficient for typical integration times of several minutes in night-time
astronomy, solar spectroscopy usually deals with exposure times well below one second
even at very high spectral resolution. The use of iodine cells is limited to the spectral range
between 500 and 630\,nm. The dense I\textsubscript{2} spectrum blends with solar lines, and deconvolution
techniques must be employed to recover the true solar spectrum from the measured data
\citep{KochWoehl1984}. While both Th-Ar and I\textsubscript{2} calibrators provide very stable references
with a reasonable number of calibration lines, a major drawback is the huge variation in
intensity of their emission/absorption lines. Also, their atomic parameters are not known
precisely enough for absolute calibration at a few meters per second.

Recently, several frequency comb-based calibration systems were proposed for astro-
nomical spectrographs  \citep[e.g.][]{Murphy2007,Braje2008,Lietal2008}, and a
few test setups readily demonstrated their superior performance for wavelength calibration
\citep{steinmetz2008,Benedick2010,wilken2010}.  \citet{wilken2010}
also showed that with grating spectrographs it is not sufficient to rely on a few calibration
lines only, as aberrations of the spectrograph optics itself or even inter-pixel irregularities of
the CCD could introduce deviations from the best-fit pixel-to-frequency curve that exceed
several 10\,m\,s\textsuperscript{-1}.

\section{The VTT Wavelength Calibration System}

The feasibility of calibrating astronomical spectrographs with frequency combs was demonstrated for the first time by \citet{steinmetz2008} using the \textit{Vacuum Tower Telescope} (VTT)
echelle spectrograph with a frequency comb operated in the infrared range. During the last
two years, a laser frequency comb-based wavelength calibration system was developed for
the VTT spectrograph in a cooperation between the Kiepenheuer-Institut f\"ur Sonnenphysik,
Freiburg, Germany and the Max-Planck-Institut f\"ur Quantenoptik, Garching, Germany. The
system was planned to cover a spectral range of at least $530\,\pm50$\,nm in the visible with a
mode separation of about 5\,pm. The VTT spectrograph has a spectral resolution of $\frac{\lambda}{\delta\lambda} > 10^6$
(0.5\,pm @ 500\,nm). For our application the rather large mode separation is required, because it easily allows us to unambiguously identify the frequencies of the comb lines by
comparison with a nearby line in the solar spectrum.

For the details of frequency comb generation and their application in astronomy we refer
to \citet{Holzwarth2000,Murphy2007,steinmetz2008} and \citet{wilken2010diss}.
Here we merely describe the basic principle of our astro-comb setup and its integration with
the spectrograph.

\subsection{Optical Frequency Combs}

The spectrum of a mode-locked femtosecond laser forms a comb-like structure
of equidistantly spaced modes with a frequency separation that is equal to
the repetition rate $\frep = 1/T$ of the laser, where $T$ is the round-trip
time of the pulse in the laser resonator. Inequality of group and phase
velocity in the resonator introduces a nonzero offset frequency
$\left|\fceo\right| < \frep$ of the laser modes so that the frequency of the
$n$th mode,
\begin{equation}
  \label{eq:comb-modefreq}
  f_n = \fceo + n \frep,
\end{equation}
is completely defined by the two radio frequencies (RFs) \fceo{} and
\frep. Equation~(\ref{eq:comb-modefreq}) establishes a direct link between
the RFs and optical spectrum. Both frequencies can be measured and controlled
very precisely, enabling the generation of optical frequencies with the same
precision that previously was only available to RF electronics. While
\frep{} can be measured straightforwardly with a fast photo-diode,
determination of \fceo{} requires additional effort. One possibility is to
use an $f$:$2f$-interferometer where the red part of the LFC spectrum is
frequency doubled and the beat note with the blue part then yields
\fceo{}. This requires the comb spectrum to span at least one octave and was
first accomplished by spectral broadening of the comb in a photonic crystal
fiber (PCF) \citep{Holzwarth2000}. The equidistant spacing of the comb modes
can be derived from theory, and no deviation from the equidistancy could be
detected experimentally at a level of a few parts in 10\textsuperscript{16}.

While frequency combs are now routinely used in many fields in metrology
and spectroscopy, their application in astronomy was delayed until recently
for two reasons: 1) the fundamental repetition rate of the comb lasers
(typically of the order of 100\,MHz or approximately 0.1\,pm\,@\,500\,nm) is
much to small to be resolved even by high-resolution astronomical
spectrographs and, 2), the wavelength coverage of the comb spectrum is too
narrow to be useful for multipurpose astronomical spectroscopy.

\subsection{The VTT Astro-Comb}

There are several solutions to the problems mentioned above. The approach we
follow is to filter the the fundamental repetition rate of the frequency
comb with Fabry-P\'erot cavities (FPCs) to a higher effective repetition
rate $f_\mathrm{rep, astrocomb} = m \frep$ where $m$ is the integer ratio of
the free spectral range (FSR) of the cavity and \frep{}. The filtered
spectrum is amplified, converted to the visible spectrum, and finally
broadened in a PCF.
\begin{figure} 
  \centerline{\includegraphics[width=\textwidth,clip=]{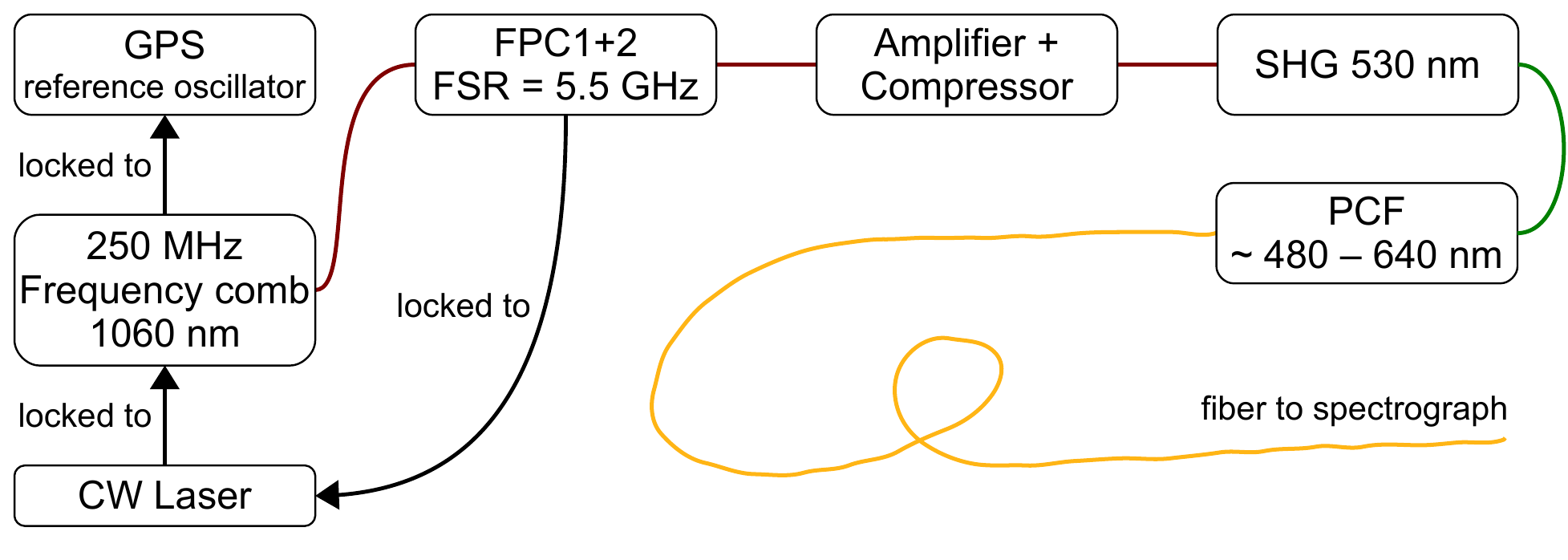}}
  \caption{Schematic view of the VTT astro-comb.}
  \label{fig:astrocombscheme}%
\end{figure}
A schematic view of the setup is shown in Figure~\ref{fig:astrocombscheme}.

The heart of the VTT astro-comb is a commercial frequency comb system (Menlo
Systems FC1000). It consists of a ytterbium fiber laser operated at 1060\,nm
with a repetition rate of 250\,MHz. The repetition rate and offset frequency
of the laser are phase-locked to a GPS disciplined reference oscillator with
a stability and accuracy of 10\textsuperscript{-12}\,@\,1\,s. While with
Ti:Sa lasers, a higher fundamental repetition rate of~$\approx 1$\,GHz would
be possible, fiber lasers have the big advantage of turnkey operation, allowing
for unattended operation by non-experts.

The comb is filtered with $m=22$ with two identical planoconcave
FPCs with a free spectral range of 5.5\,GHz and a finesse
of approximately~500. By using two cavities we obtain a suppression of the
nearest 250\,MHz sidemodes by more than 60\,dB \citep{Steinmetz2009}. The
plane mirrors of the cavities are piezo-driven for very fast adjustment of
the resonator length.  Both cavities are locked with the Pound-Drever-Hall
method \citep[e.g.][]{Black2000PDH} to the transmission signal of a
continuous wave (CW) laser with a polarization state that is orthogonal
to that of the comb laser. The CW laser itself is locked to one of the comb
modes transmitted by the cavities. Since the CW laser is comparably stable,
the cavities stay locked even when the seed laser is switched off or blocked
for a short time. As the cavities might drift out of the piezo range due to
changes in pressure or temperature, a coarse tracking system with
thermoelectrial heaters is implemented. To compensate for losses of the
laser power in the filter cavities, the signal from the seed laser is
amplified with core-pumped fiber amplifiers before passing the
cavities. After filtering, a high-power double-clad fiber amplifier and a
pulse compressor are inserted in the beam as the second-harmonic generation
(SHG) requires high pulse powers to be efficient and enough power must also
be available afterwards for spectral broadening in the PCF. We achieve good
spectral broadening with an average power of about 1.5\,W before the SHG
stage.

Up to the pulse compressor, all optical components are fiber coupled except
for a free-space section in the FPCs. This makes the
optical alignment of the system very robust; readjustment will probably
only be necessary during the regular telescope maintenance periods.

To enhance the non-linear effects that enable the spectral
broadening, the PCF is tapered \citep{Stark.etal.2011}. The characteristics
of the broadening strongly depend on the fiber which can not yet be
manufactured with completely identical properties. With most of the PCFs we
have tried so far, broad spectra ranging from approximately 480 to 640\,nm could
be obtained with enough power for short exposure times that are
still limited by the flux from the sunlight.

\subsection{Integration with the Spectrograph}

In solar physics spectroscopy usually also involves imaging of features on
the solar surface. By scanning the spectrograph slit, images can be obtained
with high spatial and spectral resolution (at the cost of temporal
resolution). However, this further complicates the use of external wavelength
calibration standards, as the calibration light must be combined with the
sunlight so that identical illumination of the spectrograph grating from
both light sources is ensured. This is probably the most critical issue of
the complete setup. Night-time astronomy is different in this aspect. Stars are always unresolved point sources; the star light and calibration
light can be fed to the spectrograph in two optical fibers which provide
excellent stability and any offset between both fibers can easily be
measured.

For observations that rely on the spatial information provided by the slit
spectrograph we are currently developing a slit illumination unit that will
allow us to align the calibration light fed to the spectrograph in a
singlemode fiber with respect to the optical axis of the
telescope. Telescope and fiber pupil are imaged on a CCD camera, and the
position of both can be tracked and aligned with a precision that translates
to a few meters/second. While this would still limit our ability for absolute
calibration, the stability of the calibration itself is not affected.

To characterize the limits of the spectrograph calibration we are also
preparing an experimental all-fiber setup where integrated sunlight from a
full-disk telescope and the calibration light is fed to the spectrograph in
the same single mode fiber. By using a single mode fiber, a perfect
mode match between calibration light and sunlight can be guaranteed,
eliminating any systematic effects related to unequal illumination of the
grating with sunlight and calibration light. This setup then can also be used
for high-precision Sun-as-a-Star spectroscopy.

\subsection{Expected Performance}

The photon-noise limited uncertainty in velocity information that can be
extracted from a single line can be estimated \citep{Brault1987} by the
relation
\begin{equation}
  \sigma_v = A \frac{\mathrm{FWHM}}{\mathrm{SNR}\sqrt{n}}
\end{equation}
where $A$ is a weighting factor that depends on the line shape (0.41 for a
gaussian according to \citet{Murphy2007}), FWHM is the full width at half
maximum of the line in meters/second, SNR is the peak signal-to-noise-ratio of the line profile, and $n$ is the number of pixels per FWHM at which the spectrum is sampled on
the CCD. For the VTT spectrograph $\sigma_v$ corresponds to 0.7\,meters/second at a
moderate SNR of 100.

\begin{figure} 
  \centerline{\includegraphics[bb=108 23 950 380,width=\textwidth]{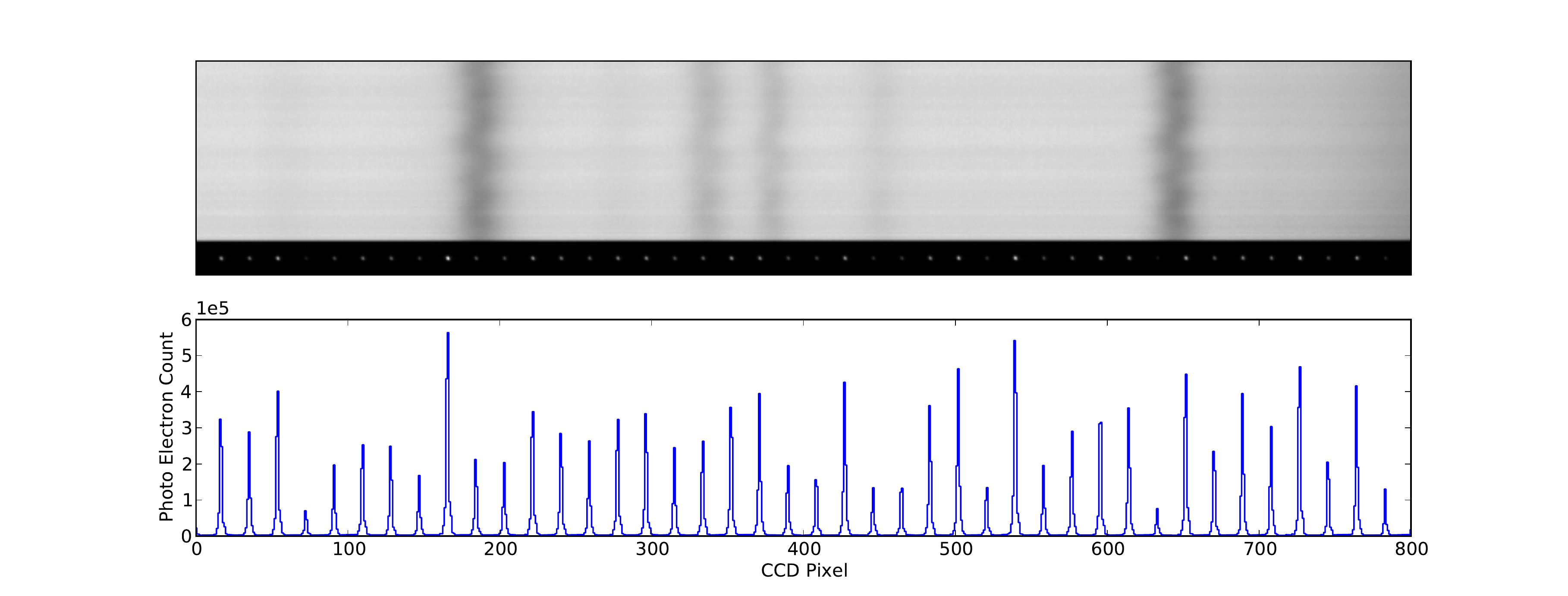}}
  \caption{A part of the solar spectrum at about 530\,nm simultaneously
    recorded with the calibration spectrum from the frequency comb on 8 Oct
    2011 17:35 UTC near the disk center (upper panel). The lower panel shows
    a cut through the LFC spectrum in horizontal direction at the same image
    scale.}\label{fig:astrocombspec}%
\end{figure}

As the VTT spectrograph is neither temperature nor pressure stabilized,
drifts of the order of 0.1\,pm (50\,m/s) per hour are common. While this
drift seems rather large, at exposure times of one second it is still below
the detection limit and can easily be tracked when the calibration spectrum
is recorded simultaneously with the science
data. Figure~\ref{fig:astrocombspec} shows a test exposure taken during the
installation campaign in October 2011. In the upper panel, a part of the
solar spectrum at about 530\,nm is shown with the calibration spectrum from
the LFC at the lower edge of the CCD image. At 530\,nm, the PCF introduces
strong intensity variations that however quickly flatten with departure from
the central wavelength of the frequency-doubled laser comb.

\section{Conclusion and Outlook}

A frequency comb-based calibration system was developed for the VTT
spectrograph and successfully installed at the telescope in October 2011.
The astro-comb allows for absolute calibration of solar spectroscopic
measurements in a broad wavelength range between approximately 480 and
640\,nm. As many spectroscopic measurements in solar physics suffer from
unreliable calibration, we see great potential for a very high-resolution
spectrograph in combination with an absolute wavelength calibration system.
The system is currently being tested and characterized thoroughly; the first
scientific campaigns are scheduled for the next observing season. We also
plan to make the system available for regular observation campaigns
beginning with the 2013 observing season.

\vspace{1em}
\noindent
{\small\textbf{Acknowledgements} This project is in part funded by the Leibniz-Gemeinschaft within the ``Pakt f\"ur Forschung und Innovation''.  A. Fischer, K. Gerber, T. Sonner, and M. Weisssch\"adel provided invaluable technical support for the setup at the VTT.}

\footnotesize
\setlength{\bibsep}{0.0pt}
\bibliographystyle{spr-mp-sola-cnd} 
\bibliography{references}  

\end{document}